\documentclass[
reprint,
superscriptaddress,
 amsmath,amssymb,
 aps,
 prl,
]{revtex4-2}

\usepackage{graphicx}
\usepackage{dcolumn}
\usepackage{bm}

\usepackage{soul}
\usepackage{xcolor}
\usepackage[colorlinks,
linkcolor=blue,
anchorcolor=blue,
citecolor=blue
]{hyperref}

\begin{document}

\preprint{APS/123-QED}

\title{Atomistic Origin of Diverse Charge Density Wave States in CsV$_3$Sb$_5$}

\author{Binhua Zhang}
\affiliation{Key Laboratory of Computational Physical Sciences (Ministry of Education), Institute of Computational Physical Sciences, State Key Laboratory of Surface Physics, and Department of Physics, Fudan University, Shanghai 200433, China.}
\affiliation{Shanghai Qi Zhi Institute, Shanghai 200030, China}

\author{Hengxin Tan}
\affiliation{Department of Condensed Matter Physics, Weizmann Institute of Science, Rehovot 7610001, Israel}

\author{Binghai Yan}
\affiliation{Department of Condensed Matter Physics, Weizmann Institute of Science, Rehovot 7610001, Israel}

\author{Changsong Xu}
\email{csxu@fudan.edu.cn}
\affiliation{Key Laboratory of Computational Physical Sciences (Ministry of Education), Institute of Computational Physical Sciences, State Key Laboratory of Surface Physics, and Department of Physics, Fudan University, Shanghai 200433, China.}
\affiliation{Shanghai Qi Zhi Institute, Shanghai 200030, China}

\author{Hongjun Xiang}
\email{hxiang@fudan.edu.cn}
\affiliation{Key Laboratory of Computational Physical Sciences (Ministry of Education), Institute of Computational Physical Sciences, State Key Laboratory of Surface Physics, and Department of Physics, Fudan University, Shanghai 200433, China.}
\affiliation{Shanghai Qi Zhi Institute, Shanghai 200030, China}

% abstract
\begin{abstract}
Kagome metals AV$_3$Sb$_5$ (A = K, Rb, or Cs) exhibit intriguing charge density wave (CDW) instabilities, which interplay with superconductivity and band topology. 
However, despite firm observations, the atomistic origins of the CDW phases, as well as hidden instabilities, remain elusive. Here, we adopt our newly developed symmetry-adapted cluster expansion method to construct a first-principles-based effective Hamiltonian of CsV$_3$Sb$_5$, which not only reproduces the established inverse star of David (ISD) phase, but also predict a series of \emph{D}$_{3h}$-\emph{n} states under mild tensile strains.  With such atomistic Hamiltonians, the microscopic origins of different CDW states are revealed as the competition of the second-nearest neighbor V-V pairs versus the first-nearest neighbor V-V and V-Sb couplings. 
Interestingly, the effective Hamiltonians also reveal the existence of ionic Dzyaloshinskii–Moriya interaction in the high-symmetry phase of CsV$_3$Sb$_5$ and drives the formation of non-collinear CDW patterns.
Our work thus not only deepens the understanding of the CDW formation in AV$_3$Sb$_5$,  but also demonstrates that the effective Hamiltonian is a suitable approach for investigating CDW mechanisms, which can be extended to various CDW systems.

\end{abstract}

\maketitle

The kagome lattice provides a fertile playground to explore intriguing correlated phenomena \cite{shores2005structurally,liu2018giant,isakov2006hard,ko2009doped}. Recently, the vanadium-based kagome metals $A$V$_3$Sb$_5$ ($A$ = K, Rb, or Cs) \cite{ortiz2019new} was reported to feature charge density wave (CDW), as well as superconductivity and Z$_2$ topology \cite{ortiz2020cs,ortiz2021superconductivity}. The interplay between CDW and other properties triggered enormous research interests \cite{zhao2021cascade,chen2021roton,jiang2021unconventional}.

\begin{figure}[t]
\centering
\includegraphics[width=8cm]{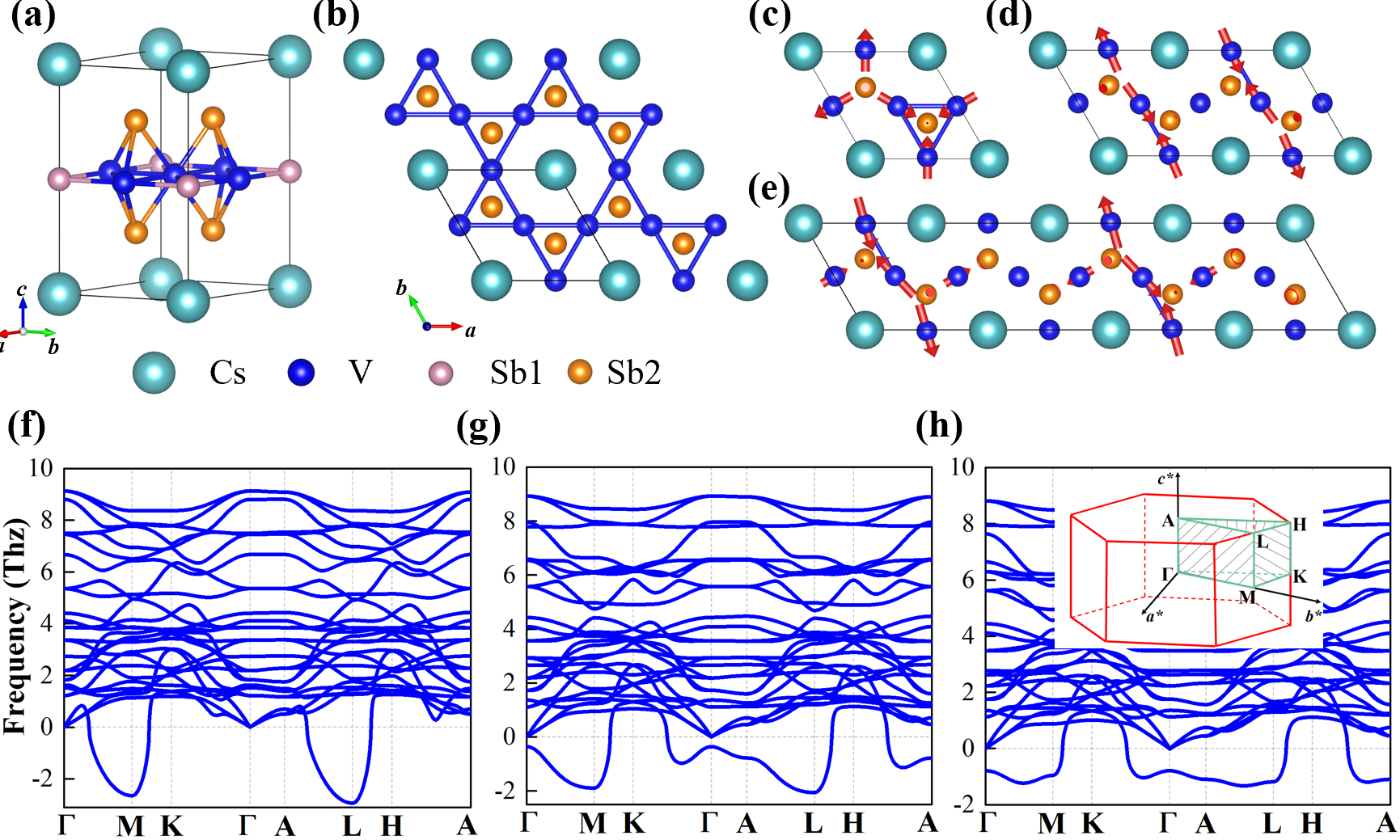}
\caption{The pristine crystal structure of CsV$_3$Sb$_5$ from (a) three-dimensional view and (b) top view. The phonon dispersion of pristine CsV$_3$Sb$_5$ (f) at 0$\%$, (g) 2$\%$, and (h) 2.8$\%$ tensile strain. The inset in (h) is the Brillouin zone for this system. (c) The vibration mode at $\Gamma$ (0, 0, 0) point in (g).(d) The vibration mode at M ($\pi$, 0, 0) point in (f). (e) The vibration mode at the middle point of the $\Gamma$-M line in (h). The red arrows show the atomic displacements in each mode.}
\label{fig:phonon}
\end{figure}

As an example of $A$V$_3$Sb$_5$, CsV$_3$Sb$_5$ exhibits diverse CDW instabilities, enriching their couplings with different electronic orders \cite{ortiz2020cs,ortiz2021superconductivity,chen2021roton,zhao2021cascade,nie2022charge,feng2021chiral,jiang2021unconventional,wu2022charge,ptok2022dynamical,ortiz2021fermi}. 
The high temperature pristine phase of CsV$_3$Sb$_5$ crystallizes in the layered structure of V-Sb sheets intercalated by Cs atoms with the space group of \emph{P\textup{6}/mmm} \cite{ortiz2019new}, as depicted in Fig.\ref{fig:phonon}(a, b). In the kagome layer, the V atoms form a kagome net interspersed by Sb1 atoms and sandwiched by two honeycomb Sb2 layers. 
CsV$_3$Sb$_5$ exhibits a CDW transition at \emph{T}$_{\rm CO}$ $\approx$ 94 K, corresponding to a 2$\times$2$\times$1 inverse star of David (ISD) pattern with $D_{6h}$ point group \cite{ortiz2020cs,tan2021charge}. At an intermediate temperature of \emph{T}$_{\rm SO}$ $\approx$ 50-60 K, a unidirectional 4\emph{a}$_0$ or 5\emph{a}$_0$ stripe order has been observed on Sb-terminated surface, the atomic pattern of which is still elusive \cite{zhao2021cascade,liang2021three,nie2022charge,wang2021electronic,ye2022structural,zheng2022emergent}. 
With a slightly lower temperature of \emph{T}$_{\rm N}$ $\approx$ 40 K, a nematic state occurs, reducing the rotation symmetry from \emph{C}$_6$ to \emph{C}$_2$ \cite{nie2022charge}. Also, the 2$\times$2$\times$2 and 2$\times$2$\times$4 CDW phases have been observed, associated with the stacking of the adjacent kagome layers \cite{PhysRevB.105.195136,PhysRevMaterials.7.024806}. Moreover, such CDW instabilities compete with each other and the ground state varies with doping, surfacing and pressure \cite{du2021pressure,chen2021double,yu2021unusual,oey2022fermi}.
Among different CDW states, the origin of ISD state was successfully explained with soft phonons in a collective manner \cite{tan2021charge}. However, it still lacks an overall understanding of the known and possible hidden CDW instabilities in  CsV$_3$Sb$_5$, especially in an atomistic perspective.

In this work, we apply the first-principles-based effective Hamiltonian approach \cite{SM,lou2021pasp}, which is widely used in ferroelectric systems \cite{zhong1995first,bellaiche2000finite}, to investigate the atomic CDW instabilities in CsV$_3$Sb$_5$. 
Significantly, a series of breathing-kagome-like \emph{D}$_{3h}$ phases are predicted with mild tensile epitaxial strain. The atomic origins for the new  \emph{D}$_{3h}$ phases and the known ISD state are comprehensively understood in terms of V-V and V-Sb pairs.
Particularly, the atomic Hamiltonian reveals the existence of ionic Dzyaloshinskii–Moriya interaction (DMI), which is found to be the driving force of the formation of these nonlinear CDW patterns.

\begin{figure}[t]
    \centering
    \includegraphics[width=8cm]{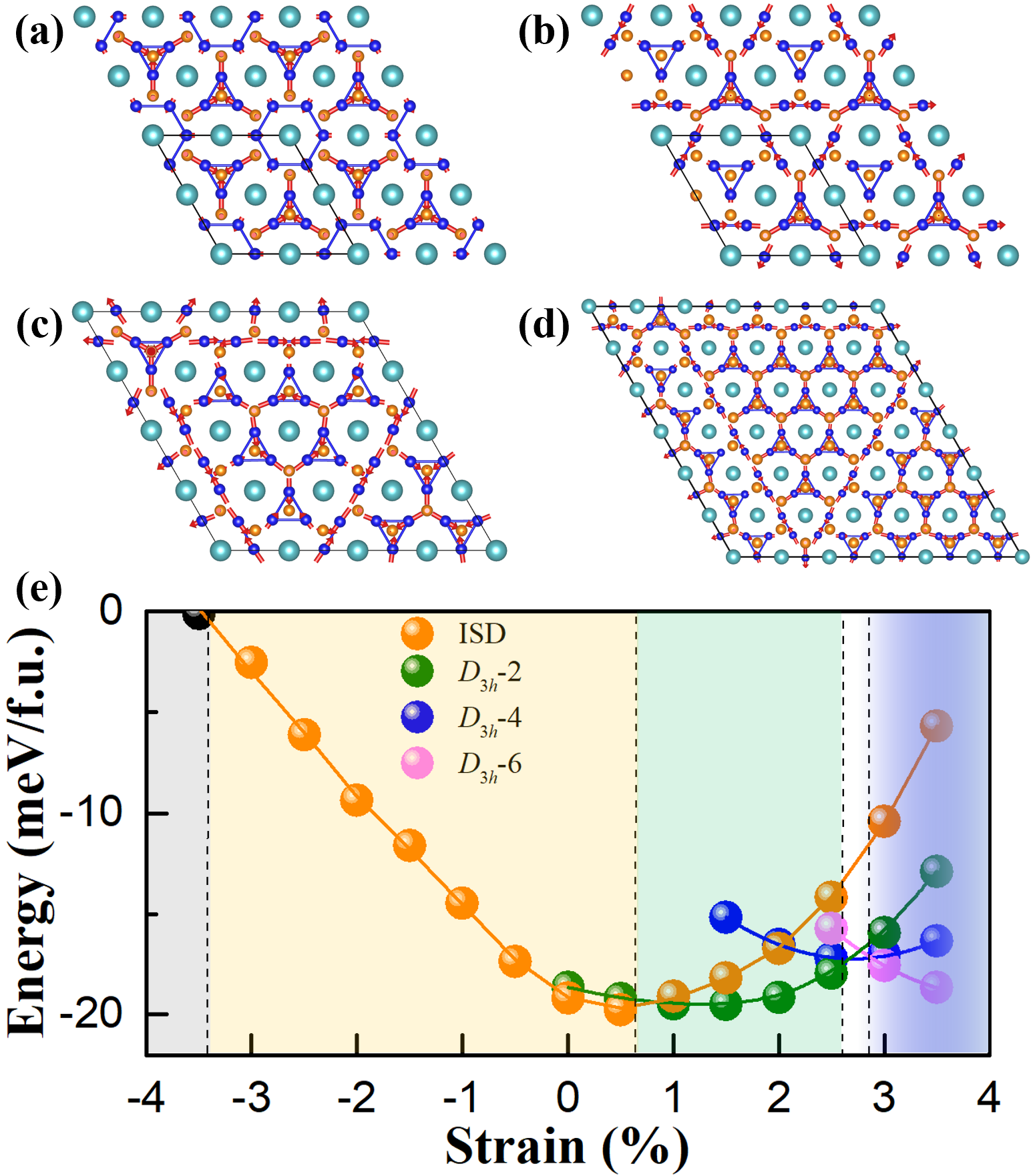}
    \caption{The CDW states of CsV$_3$Sb$_5$, (a) ISD, (b) \emph{D}$_{3h}$-2, (c) \emph{D}$_{3h}$-4, and (d) \emph{D}$_{3h}$-6. Black solid lines mark the unit cell in the distorted CDW phase. The red arrows show the atomic displacements. (e) The strain dependence of total energy of these CDW states with respect to the pristine phase, in unit of meV per formula unit (f.u.) The black ball in panel (e) represents the pristine phase. }
    \label{fig:dft}
\end{figure}

{\it CDW transition from DFT.} 
Phonon dispersion of the pristine phase at zero strain is displayed in Fig. \ref{fig:phonon}(f). The imaginary phonon mode at M point [e.g., $2\pi$(1/2, 0, 0)] indicates a 2$\times$1$\times$1 modulation of the structure, which mainly involves the V-V in-plane movements as shown in Fig. \ref{fig:phonon}(d). A linear combination of the soft modes at the three symmetry-equivalent M points give rise to the star of David (SD) or ISD [Fig. \ref{fig:dft}(a)] pattern within a 2$\times$2$\times$1 supercell. Note that the ISD structure exhibits inverse atomic displacements compared to the SD structure. The ISD phase is energetically favored (see Fig. S1 in Supplemental materials (SM) \cite{SM}), which is consistent with previous results \cite{tan2021charge} and will be explained in terms of the microscopic interactions discussed below. Similarly, the three L-point soft modes indicate a 2$\times$2$\times$2 structural reconstruction, which is manifested by the ISD + ISD stacking pattern with $\pi$-phase shift (see Fig. S2 \cite{SM}), agreeing well with previous studies \cite{tan2021charge,liang2021three}. For simplicity, we will focus on the 2$\times$2$\times$1 CDW orders in the following discussion unless stated otherwise.

Let us then look at the strain effects. Recent experiments report different critical temperatures between bulk and nano flakes of CsV$_3$Sb$_5$, indicating surface strain releasing plays an important role \cite{song2021competition,song2021competing2,zhang2022emergence}. We thus investigate the phonons of the CsV$_3$Sb$_5$ pristine phase with in-plane biaxial strain (see Method in Part I of SM \cite{SM}). As shown in Fig. \ref{fig:phonon}(g, h), the phonon instability changes rapidly with increasing tensile strain. An additional imaginary phonon mode appears at $\Gamma$ point at 2\% strain and is found to correspond to a breathing kagome mode, in which the V atoms move inward in one triangle and outward in the adjacent triangles [Fig. \ref{fig:phonon}(c)]. Superposing this mode onto the ISD structure leads to a new 2$\times$2$\times$1 structural reconstruction, characterized by shrinked triangle of V atoms with surrounding V-V dimers (\emph{D}$_{3h}$-2 phase in Fig. \ref{fig:dft}(b), here, the $n \times n$ supercell with \emph{D}$_{3h}$ symmetry is denoted by \emph{D}$_{3h}$-\emph{n}, [see Part VIII-X, XIV of SM for band structures, scanning tunneling microscopy (STM), structural information and XRD patterns of these states \cite{SM}]. In addition, the lowest imaginary frequency moves from M point under zero strain to approximately the middle of $\Gamma$-M for 2.8$\%$ tensile strain, corresponding to the mode characterized by the V-V in-plane dimerization within a 4$\times$1$\times$1 supercell [Fig. \ref{fig:phonon}(e)]. 
Such dimerization with a phase shift along ${\bm a}$ 
axis strongly implies a new possible 4\emph{a}$_0$ stripe order \cite{zhao2021cascade,liang2021three,nie2022charge,zheng2022emergent,ye2022structural}, consistent with the observation of Ref. \cite{zhao2021cascade} (see Part IX for STM images \cite{SM}).
Considering the multiplicity of above imaginary mode, the new \emph{D}$_{3h}$-4 state [Fig. \ref{fig:dft}(c)] can be constructed. Interestingly, as strain goes more tensile, it is found that the lowest frequency moves more and more toward $\Gamma$ point, indicating larger and larger supercells. For example, the \emph{D}$_{3h}$-6 state [Fig. \ref{fig:dft}(d)] can emerge at the strain of 3.5\% (Fig. S4 \cite{SM}). Such results thus imply a continuum evolution of CDW states with increasing strain.

%{\it DFT phase diagram.} 
In order to determine the ground states at different strains, we perform density functional theory (DFT) calculations on various CDW configurations, which are condensated from the soft phonons (see Part V for all considered structures \cite{SM}). 
Figure \ref{fig:dft}(e) shows the energy-vs-strain curves of selected low-energy states.
It indicates that, for the compressive strains exceeding -3.4$\%$, CsV$_3$Sb$_5$ prefers the pristine phase [Fig.\ref{fig:phonon}(a, b)], which is in line with the observation that pristine phase becomes ground state under hydrostatic pressure \cite{zheng2022emergent}. From -3.4$\%$ compressive strain to 0.7$\%$ tensile strain, the ISD state shown in Fig. \ref{fig:dft}(a) is energetically favorable, which is also consistent with observations \cite{ortiz2020cs}. 
Particularly, between tensile strain of  0.7$\%$ and 2.7\%, the aforementioned new \emph{D}$_{3h}$-2 CDW state is stabilized (see Fig. S8 for phonon dispersion and XIII for the prediction of CsV$_3$Bi$_5$ which exhibits \emph{D}$_{3h}$-2 phase in the absence of any strain \cite{SM}). As the tensile strain increases, the V-V dimers form larger and larger triangle, which encircles more and more shrinked triangles of V atoms, corresponding to the \emph{D}$_{3h}$-4 phase [Fig. \ref{fig:dft}(c)], \emph{D}$_{3h}$-6 phase [Fig. \ref{fig:dft}(d)], etc. It is thus reasonable to  speculate that, for even stronger tensile strain, the ground state will transform to a phase with only shrinked triangles, which is actually the breathing kagome state. Note that the \emph{D}$_{3h}$-\emph{n} phases also prefer that adjacent kagome layers have a $\pi$-shift. Such  $\pi$-shift further lowers  the rotation symmetry from \emph{C}$_3$ to \emph{C}$_2$ (Fig. S3 \cite{SM}).

\begin{table}[t]
\caption{\label{tab:eff}%
Selected invariants and parameters of the effective Hamiltonian of CsV$_3$Sb$_5$. The parameter $k_{\alpha, n}$ and $J_{\alpha-\beta,n}$ represent the $n^{\rm{th}}$-order coefficient of the local-mode self-energy of $\alpha$ atom and the 1NN cluster interactions of $\alpha$-$\beta$ ($\alpha$/$\beta$= V/Sb2) pairs, respectively. Here $k'_{\alpha, n}$ and $J'_{\alpha-\beta, n}$ are used for the coefficients of different invariants for one atom and the 2NN cluster interactions. The values of these coefficients (in meV/Bohr$^2$) are shown for 0\% and 2\% strains. The invariants consist of $u_i$, which represents the local soft mode at site $i$. \{XYZ\} represents the local basis of each atom. The cluster associated with each invariant is shown in Fig. \ref{fig:eff}(a-c). The ${\textbf D}_{12}$$\cdot$({\textbf U}$_1$$\times${\textbf U}$_2$) in the last row represents the 1NN i-DMI, corresponding to Fig. \ref{fig:DMI}. The positive and negative signs of $D^z_{12}$ (the scalar value of ${\textbf D}_{12}$) indicate the +z and -z direction, respectively. {\textbf U}$_i$ represents the displacement vector of i$^{th}$ V atom in glabal basis. The lengths between atoms of each pair are listed in the second column and the values in parentheses correspond to the length under 2$\%$ tensile strain.
}
\begin{ruledtabular}
\begin{tabular}{c c c c c}
\textrm{Coefficient}&
\textrm{Length (\AA)}&
\textrm{Invariant}&
\textrm{0\%}&
\textrm{2\%}\\
\colrule

$k_{\textup{V,2}}$ &	 -   & 	\emph{u}$_{1X}^2$ &	1201 &	883\\
$k'_{\textup{V,2}}$ &  - &		\emph{u}$_{1Y}^2$ &	613 &	532\\
$k_{\textup{V,4}}$	& - &	\emph{u}$_{1Y}^4$ &	432 &	616\\
$k_{\textup{Sb2,2}}$ & - &		\emph{u}$_{1Z}^2$ &	1238 &	1271\\
$J_{\textup{V-Sb2,2}}$ &	2.75 (2.74) &	\emph{u}$_{1Y}$\emph{u}$_{1Z}$ &	498 &	540\\
$J_{\textup{V-V,2}}$	& 2.72 (2.77) &	\emph{u}$_{1Y}$\emph{u}$_{2Y}$ &	142 &	240\\
$J'_{\textup{V-V,2}}$ &	4.71 (4.81) &	\emph{u}$_{1Y}$\emph{u}$_{2Y}$ &	-150 &	-160\\
$J'_{\textup{V-V,3}}$ &	4.71 (4.81) &	\emph{u}$_{1Y}^2$\emph{u}$_{2Y}$+ \emph{u}$_{1Y}$\emph{u}$_{2Y}^2$ &	-250 & -296\\ \hline
$D_{12}^z$	& 2.72 (2.77) &	 ${\textbf D}_{12}$$\cdot$({\textbf U}$_1$$\times${\textbf U}$_2$) &	158 &	-39\\
\end{tabular}
\end{ruledtabular}
\end{table}

\begin{figure}[t]
    \centering
    \includegraphics[width=8cm]{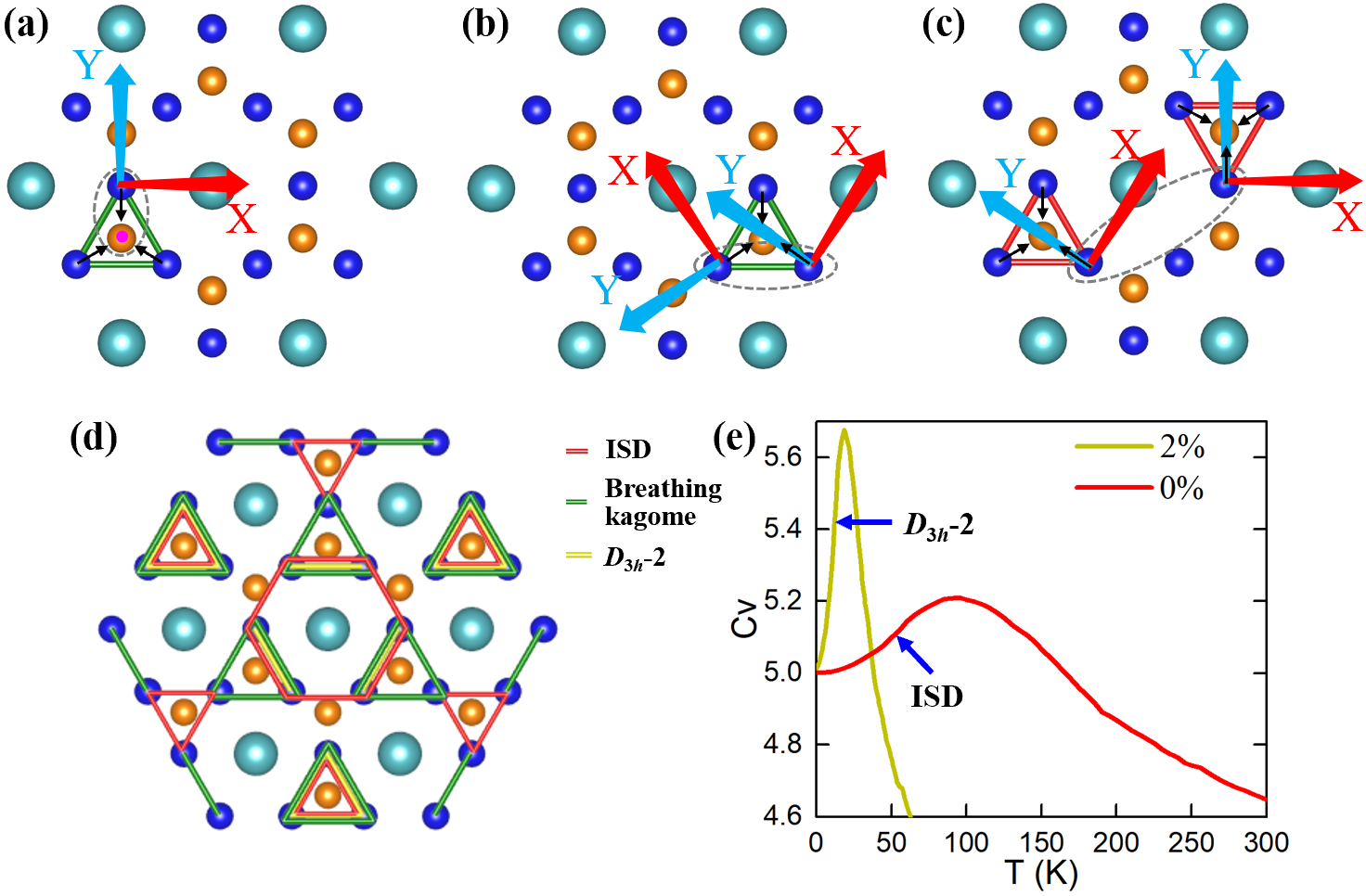}
    \caption{The effective Hamiltonian investigation and Monte Carlo simulations of CsV$_3$Sb$_5$. (a) V-Sb2 pairs. (b) The 1NN V-V pairs. (c) The 2NN V-V pairs. The local \{XY\} basis of the V atom is indicated by red and blue arrows. The local \{Z\} basis of the Sb2 atom (above kagome layer) is along out-of-plane direction, indicated by the magenta point.  The green and red bonds represent the breathing kagome and ISD deformation, corresponding to Fig. \ref{fig:phonon}(c) and Fig. \ref{fig:dft}(a), respectively. The gray dashed circles are used to mark the V-Sb2 and V-V pairs. The black arrows show the displacements driven by the pair interaction. (d) The superposition of breathing kagome and ISD state. The breathing kagome, ISD and \emph{D}$_{3h}$-2 deformation are represented by green, red, yellow bonds, respectively. (e) Temperature evolution of specific heat at 0$\%$ and 2$\%$ strain.}
    \label{fig:eff}
\end{figure}

{\it Atomic driving forces from Effective Hamiltonian.} 
In order to get more insights into the microscopic origins of the CDW states, we develop first-principles-based effective Hamiltonians for CsV$_3$Sb$_5$ (see Table. S5 for KV$_3$Sb$_5$ and RbV$_3$Sb$_5$). 
Here we adopt the symmetry-adapted cluster expansion method, as implemented in the PASP software \cite{lou2021pasp}. Such method enables defining local modes ${\bm u}$ (atomic displacements) and generating all possible forms of interactions, i.e., the invariants. With input energies of random structures from DFT, the coefficients of such invariants can be fitted using machine learning method for constructing Hamiltonian \cite{li2020constructing}. Note that the effect of electron-phonon coupling is implicitly included in these coefficients.
For the present case of CsV$_3$Sb$_5$, local modes are defined as (i) in-plane displacements of V atoms and (ii) out-of-plane displacements of Sb2 atoms, while other minor movements are neglected.
The initial model contains enough invariants, including the  self-energy of local mode up to fourth-order and the cluster interactions  up to the four-body and fourth-order. 
After repeated fitting and refining, it arrives at below Hamiltonian,
\begin{equation}
\mathcal{H}= \varepsilon^{\rm self}(\{{\bm u}_i\})+\varepsilon^{\rm pair}(\{{\bm u}_i,{\bm u}_j\})
\end{equation}
where $\varepsilon^{\rm self}$ is the single site energy and $\varepsilon^{\rm pair}$ is the energy of pair interaction, which indicates that three-body and four-body interactions are not important here. Herein the two effective Hamiltonians are constructed separately for strains of 0\% and 2\% (see more strains in Table. S3), with the primary coefficients listed in Table \ref{tab:eff}, while a full model and details can be found in Part VII in SM \cite{SM}. The energies from DFT and effective Hamiltonians are in good consistency (see Fig. S11).
Note that the forms of interactions are expressed in local basis [see Fig. \ref{fig:eff}(a-c)] for simplicity.
As shown in Fig. \ref{fig:eff}(e), Monte Carlo simulations with the obtained effective Hamiltonians indicate that (i) at zero strain, ISD state forms below 100 K, which is rather close to the measured \emph{T}$_{\textup{CO}}$ = 94 K \cite{ortiz2020cs}; and (ii) at 2\% strain, the \emph{D}$_{3h}$-2 state emerges below \emph{T}$_\textup{D}$ = 20 K, indicating that the tensile strain largely lowers the CDW critical temperature. 

We now work on understanding the formation of ISD and \emph{D}$_{3h}$-2 states by analysing the invariants and their coefficients. According to the experience of  Hamiltonians for ferroelectrics, the second-order interactions usually act as the driving force of the phase transition, while that the fourth-order interactions prevent too large atomic displacements \cite{zhong1995first}. Such mechanisms lead to the well-known double-well energy landscape.
Here, as shown in Table \ref{tab:eff}, single site terms, either second or fourth orders, all have positive coefficients, indicating that neither ISD nor \emph{D}$_{3h}$-2 states are initiated by on-site instability.
On the other hand, three second-order pair interactions are found to drive the deformations to CDW states (note that the local basis \{XYZ\} is adopted): (i) the first-nearest neighbor (1NN) V-Sb2 interaction, with the form of $u_{1Y}u_{2Z}$ and a positive coefficient, drives V atoms toward in-plane trimerization (shrinked triangle) and Sb2 atoms moving toward out-of-plane direction [Fig. \ref{fig:eff}(a)]; (ii) the 1NN V-V interaction, with the form of $u_{1Y}u_{2Y}$ and also a positive coefficient, also tends to form trimers of V atoms [Fig. \ref{fig:eff}(b)]; and (iii) the second-nearest neighbor (2NN) V-V interaction, with the form of $u_{1Y}u_{2Y}$ and a negative coefficient, drives trimerization of V atoms at a large distance [Fig. \ref{fig:eff}(c)] (the sign of the coefficient of these pair interactions can be related to the band splitting near Fermi level, which may lower the energy relative to the band crossing at the inverse sign, see Figs. S13-14 \cite{SM}). 
Note that (i) and (ii) lead to trimers with the same orientation, while (iii) results in trimers with opposite orientations.
Such analyses indicate that (i) and (ii) tend to stabilize breathing kagome state, while (iii) favors ISD state.
Interestingly, the 2NN V-V pair has a third-order interaction of \emph{u}$_{1Y}^2$\emph{u}$_{2Y}$+ \emph{u}$_{1Y}$\emph{u}$_{2Y}^2$, which naturally lifts the energy degeneracy between SD and ISD, or between  \emph{D}$_{3h}$-\emph{n} and inverse \emph{D}$_{3h}$-\emph{n}. We then look at the strain effects. At zero strain, the 2NN V-V pairs play the dominant role, which results in the ISD deformation. When a 2$\%$ tensile strain is applied, values of $J_{\textup{V-Sb2,2}}$ and 1NN $J_{\textup{V-V,2}}$ are significantly enhanced (Table \ref{tab:eff}), which tends to stabilize the breathing kagome pattern. This is understandable since the larger lengths of 1NN V-V pairs under tensile strain enhance the tendency close to each other. Meanwhile, the 2NN $J'_{\textup{V-V,2}}$ barely changes with strain and thus still favors the formation of ISD phase. The interplay of such two strain effects is shown in Fig. \ref{fig:eff}(d), which clearly demonstrate that the \emph{D}$_{3h}$-2 reconstruction (yellow bonds) originates from superposing breathing kagome (green bonds) and the ISD (red bonds) deformations (see Part IV for the superposition of \emph{D}$_{3h}$-4 and \emph{D}$_{3h}$-6 states).

\begin{figure}[t]
    \centering
    \includegraphics[width=8cm]{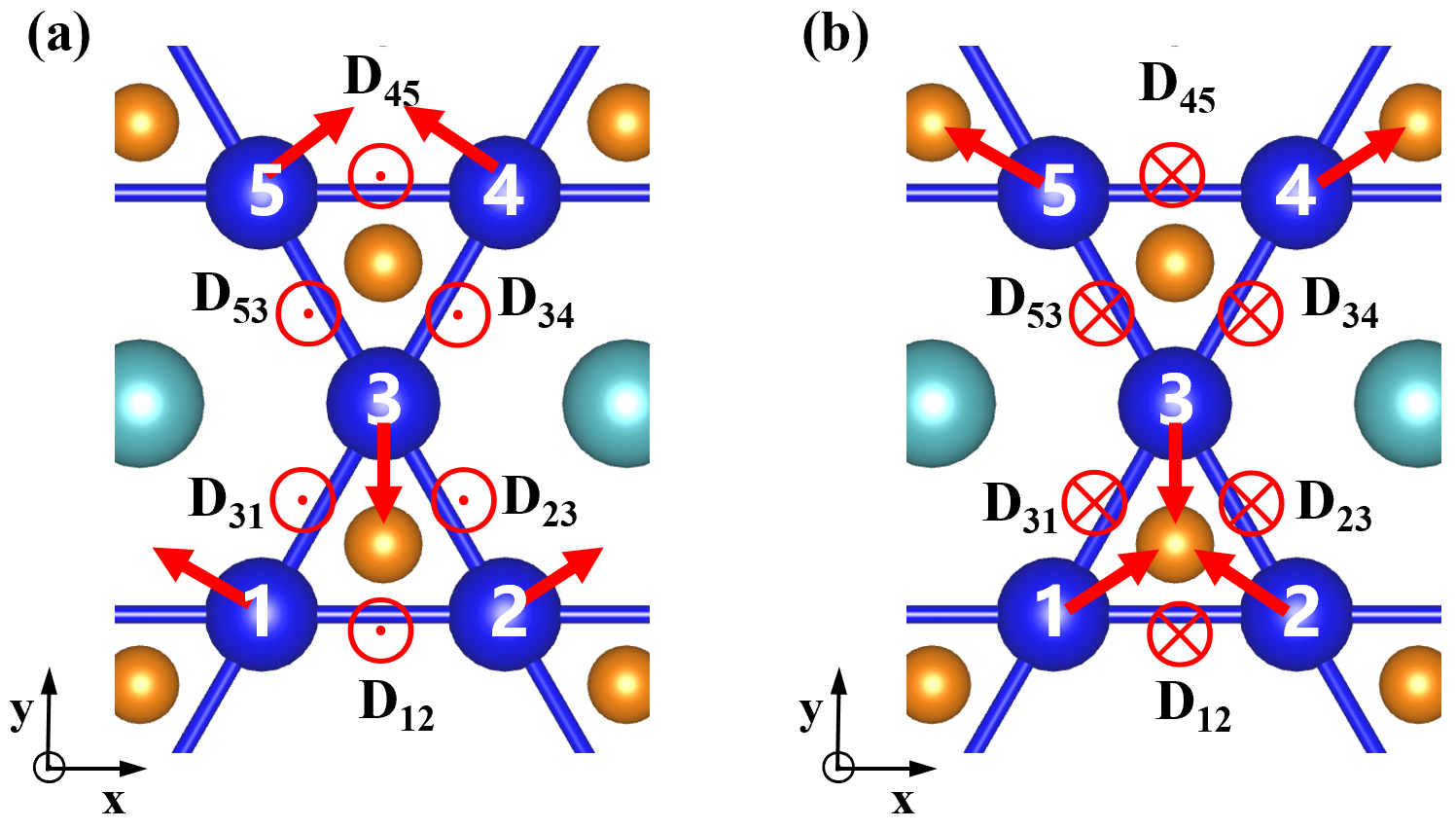}
    \caption{The sketch for the 1NN i-DMI in CsV$_3$Sb$_5$ and possible distortion modes of V atoms (blue balls). The red arrows show the atomic displacements U. By symmetry analyses, the marked ${\textbf D}_{12}$, ${\textbf D}_{23}$, ${\textbf D}_{31}$, ${\textbf D}_{31}$, ${\textbf D}_{34}$, ${\textbf D}_{45}$, and ${\textbf D}_{53}$ have the same direction and magnitude. Note that ${\textbf D}_{12}$ and ${\textbf D}_{21}$ have opposite sign. When ${\textbf D}_{12}$ is along the z direction, it tends to form (a) the ISD-like pattern (corresponding to the distortion of hexagons and adjacent atoms in Fig. \ref{fig:dft}(a)). Otherwise, (b) the breathing kagome pattern [see Fig. \ref{fig:phonon}(c) and Fig. \ref{fig:dft}(b)-(d)] will be stabilized. Note that the global $\{xyz\}$ basis is adopted here and $\otimes$ and $\odot$ represent the -z and +z direction, respectively.}
    \label{fig:DMI}
\end{figure}

Let us now further understand the nature of V-V interactions in the global basis.
As shown in Fig. \ref{fig:DMI}, between 1NN V-V pairs, e.g., the pair of V1-V2, our Hamiltonians indicate an interaction in the form of U$_{1x}$U$_{2y}$-U$_{1y}$U$_{2x}$, which can be written as ${\textbf D}_{12}$$\cdot$({\textbf U}$_1$$\times${\textbf U}$_2$), where ${\textbf D}_{12}$ has only $z$ component and the displacements of ${\textbf U}_1$ and ${\textbf U}_2$ possess only in-plane components. Such interaction is analogous to the magnetic DMI and is thus named as ionic DMI (i-DMI) here. 
As {\textbf U}$_1$$\times${\textbf U}$_2$ has the same symmetry transformation properties as {\textbf S}$_1$$\times${\textbf S}$_2$ (where {\textbf S} represents spins), the direction of the DM vector in the i-DMI case is determined by the crystal structure in the same way as the magnetic DMI case. According to Moriya's rules \cite{moriya1960new}, DM vector of V1-V2 pair is along $\pm z$ direction, as a result of an in-plane mirror including V1-V2 pair and a mirror perpendicular to V1-V2 pair. Such i-DMI favors related displacements being perpendicular to each other with the $xy$ plane, which is similar to the effects of magnetic DMI. However, it finally forms the 120$^\circ$ patterns, as shown in  Fig. \ref{fig:DMI}, due to the competition associated with kagome lattice (see Part XI of SM \cite{SM}).
Note that the i-DMI of 2NN V-V pairs is not discussed here due to their negligible values. 

From the aspect of i-DMI, we now further understand the CDW transition from ISD to \emph{D}$_{3h}$ state.
At zero strain, the i-DMIs exhibit a positive value of $D^z_{12}=158$ meV, which stabilizes a ISD-like state, as shown in Fig. \ref{fig:DMI}a, with an energy gain of $E_a=-\sqrt{3}D^z_{12}|{\textbf U}_1|^2$ per V atom. For the exact ISD pattern that CsV$_3$Sb$_5$ displays, such i-DMI leads to an energy gain of $-\frac{1}{2}D^z_{12}|{\textbf U}_1|^2$,  while the pattern shown in Fig. \ref{fig:DMI}b results in an energy cost of $-E_a$  per V atom.
Note that it is assumed that all V atoms share the same amount of displacements.
In contrast, with a tensile strain of 2\%, the i-DMIs change their signs and render a negative value of $D^z_{12}=-39$ meV, which favors the formation of breathing kagome pattern, as shown in Fig. \ref{fig:DMI}b. In such case of negative $D^z_{12}$, the breathing kagome pattern yields an energy gain of $E_b=\sqrt{3}D^z_{12}|{\textbf U}_1|^2$ per V atom, while the ISD-like pattern cost energy of $-E_b$ per V atom.
It is thus clear that, above a critical point of tensile strain, the i-DM vectors flip their directions, which deforms the ISD phase and stabilizes the $D_{3h}$ breathing kagome state.

The present discovery of i-DMI as a driving force to induce CDW states provides a new perspective to understand structural phase transitions. We note that the existence of i-DMI was first  pointed out in the low energy distorted  perovskites \cite{zhao2021dzyaloshinskii}. Here, we find that CsV$_3$Sb$_5$ is the first system where the i-DMI exists in the high-symmetry high temperature structure, and is responsible for the structure phase transition to the low energy structure.
Considering that the kagome lattice intrinsically lacks inversion center between first nearest neighbors, the i-DMI is likely to be a common mechanism to induce structural distortions there.

In summary, combining DFT calculations, we apply the effective Hamiltonian approach to CsV$_3$Sb$_5$ and predict a series of \emph{D}$_{3h}$-\emph{n} CDW phases and a possible 4\emph{a}$_0$ pattern, besides the known ISD structure. 
The nature of such CDW states is revealed as that the 2NN V-V pairs dominate the formation of ISD state, while that the 1NN V-V and V-Sb2 interactions tends to stabilize the \emph{D}$_{3h}$-\emph{n} breathing kagome state. The ionic DMI is identified as a new driving force to induce structural distortions, which is expected to commonly exist in different kagome CDW systems. Experimental verification of the new atomics mechanisms and new CDW states revealed in this work is called for.

\begin{acknowledgments}
We thank Prof. Tao Wu for useful discussions on strain effects. H.T. acknowledges Shangfei Wu for helpful discussions. We acknowledge financial support from the National Key R$\&$D Program of China (No. 2022YFA1402901), NSFC (Grants No. 11825403, No. 11991061, No. 12188101, No. 12174060, and No. 12274082), the Guangdong Major Project of the Basic and Applied basic Research (Future functional materials under extreme conditions–2021B0301030005), and Shanghai Pilot Program for Basic Research—FuDan University 21TQ1400100 (23TQ017). C. X. also acknowledges support from the Shanghai Science and Technology Committee (Grant No. 23ZR1406600). B. Z. also acknowledge the support from China Postdoctoral Science Foundation (grant No. 2022M720816).
\end{acknowledgments}

%%%%% CLEAR DOUBLE PAGE!
%\newpage{\pagestyle{empty}\cleardoublepage}

\end{document}